\documentclass{article}%
\usepackage{amsfonts}
\usepackage{amsmath}
\usepackage{amssymb}
\usepackage[a4paper]{geometry}
\usepackage[pdftex]{graphicx}%
\newtheorem{theorem}{Theorem}

\newtheorem{lemma}[theorem]{Lemma}

\newtheorem{proposition}[theorem]{Proposition}

\newenvironment{proof}[1][Proof]{\noindent\textbf{#1.} }{\ \rule{0.5em}{0.5em}}
\pdfoutput=1
\begin{document}

\title{Point Interaction Hamiltonians in Bounded Domains}
\author{Ph. Blanchard\thanks{Fakult\"{a}t f\"{u}r Physik \& BiBoS, Universit\"{a}t
Bielefeld, D-33501 Bielefeld, Germany. E-mail:
blanchard@Physik.Uni-Bielefeld.DE}, R. Figari\thanks{Istituto Nazionale di
Fisica Nucleare (INFN), Sezione di Napoli and Dipartimento Scienze
Fisiche,\ Universit\`{a} degli Studi Napoli \textquotedblleft Federico
II\textquotedblright, Via Cintia, Monte S. Angelo I-80126 Napoli, Italy.
E-mail: figari@na.infn.it}, A. Mantile\thanks{Dipartimento di Matematica e
Applicazioni \textquotedblleft R. Caccioppoli\textquotedblright%
\ Universit\`{a} degli Studi Napoli \textquotedblleft Federico
II\textquotedblright, Via Cintia, Monte S. Angelo I-80126 Napoli, Italy.
Supported by a fellowship of Fak\"{u}ltat f\"{u}r Mathematik, Universit\"{a}t
Bielefeld, D-33501 Bielefeld, Germany. E-mail: andrea.mantile@dma.unina.it}}
\date{}
\maketitle

\begin{abstract}
Making use of recent techniques in the theory of selfadjoint extensions of
symmetric operators, we characterize the class of point interaction
Hamiltonians in a 3-D bounded domain with regular boundary. In the particular
case of one point interaction acting in the center of a ball, we obtain an
explicit representation of the point spectrum of the operator togheter with
the corresponding related eigenfunctions. These operators are used to build up
a model-system where the dynamics of a quantum particle depends on the state
of a quantum bit.

\end{abstract}

\section{Introduction}

The name point interactions refers to a particular class of selfadjoint
operators obtained as perturbations of the Laplacian supported by discrete
sets of points. In $\mathbb{R}^{d}$, $d\leq3$, they have been rigorously
characterized using the theory of selfadjoint extensions of symmetric
operators, \cite{fad1}, and it has been shown, \cite{Albeverio1}, that their
spectral properties (eigenvalues and eigenfunctions), and therefore all the
physical relevant quantities related to, can be explicitly computed. In this
connection, point interactions have been used to build up solvable models for
those physical systems where short range forces are supposed to play a role.
In spite of their interest in fundamental as well as in applied physics, few
results are actually available on this class of interactions in bounded
domains (the semigroup generated by a one point interaction in the case of the
heat equation in bounded domains is considered in \cite{Caspers}). Making use
of recent techniques in the theory of selfadjoint extensions, we give, in the
following, a general characterization of point interactions in bounded domains
of $\mathbb{R}^{3}$. In particular we will show that point-supported
perturbations of the Laplacian can produce a global change of its spectrum.

The results presented in this work can be used to define model-systems of
particles confined in finite regions and interacting with arrays of scattering
centers of delta type. Similar models can be used in many different
applications and have been already considered in the case of unbounded
domains. In particular, our purpose is to provide a model of a finite volume
quantum measurement apparatus whose interaction with the system under
measurement is described by a delta-shaped quantum potential. An analogous
problem in $\mathbb{R}^{1}$ and $\mathbb{R}^{3}$ have been investigated in
\cite{Caccia 1}.

In the next Section we introduce a classification of the singular
perturbations of the Laplacian by means of selfadjoint linear relations
(Theorem \ref{Theorem H_AB}). This parametrization will be used in Section 3
in order to define the selfadjoint operators associated to $N$ point
interactions acting in a 3-D bounded domain. In Section 4 one point
interaction in a ball is considered: we will show that the symmetry of the
system allows, in this case, an explicit computation of the spectral
properties of the operator. In Section 5 we define point interactions between
a quantum particle and a finite dimensional quantum system. Possible
applications of this class of models in the framework of quantum information
theory are presented.

\section{Parametrization of selfadjoint extensions}

We start recalling notation together with few definitions and results in the
theory of selfadjoint extensions of symmetric operators in the form introduced
in \cite{Gorba}.

We consider a closed densely defined symmetric operator, $H_{0}$, acting on an
Hilbert space $\mathcal{H}$, with deficiency indices $\left(  n,n\right)  $. A
triple $\left(  V,\Gamma_{1},\Gamma_{2}\right)  $ formed by an auxiliary
Hilbert space, $V$, and a couple of bounded linear operators $\Gamma
_{i=1,2}:D(H_{0}^{\ast})\rightarrow V$, such that the following conditions
hold\footnote{Here $\left(  \cdot,\cdot\right)  $ and $\left\langle
\cdot,\cdot\right\rangle $ denotes the scalar products in $L^{2}(\Omega)$ and
$V$ respectively.}:%
\begin{equation}
\left(  \psi,H_{0}^{\ast}\varphi\right)  -\left(  H_{0}^{\ast}\psi
,\varphi\right)  =\left\langle \Gamma_{1}\psi,\Gamma_{2}\varphi\right\rangle
-\left\langle \Gamma_{2}\psi,\Gamma_{1}\varphi\right\rangle \quad\forall
\psi,\varphi\in D(H_{0}^{\ast}) \label{b.v.s. 1}%
\end{equation}%
\begin{equation}
\text{the map }\left(  \Gamma_{1},\Gamma_{2}\right)  :D(H_{0}^{\ast
})\rightarrow V\oplus V\quad\text{is surjective} \label{b.v.s. 2}%
\end{equation}
defines a \emph{boundary value space }for $H_{0}$. This structure, which has
been proved to exist for any symmetric operator with equal deficiency indices
(see \cite{Gorba}, theorem 3.1.5), can be used to parametrize the selfadjoint
extensions of $H_{0}$ by means of generalized boundary conditions of the form
$A\Gamma_{1}\psi=B\Gamma_{2}\psi$, where $A$ and $B$ are bounded linear
operators on $V$. The conditions to be imposed on the operators $A$ and $B$,
in order that they may describe selfadjoint linear relations, have been given
in \cite{Panka 1} for any value of the deficiency indices $0<n\leq\infty$. In
particular, for finite values of $n$, the complex $n\times n$ matrices $A,B$
have to satisfy the properties%
\[%
\begin{array}
[c]{ll}%
\left[  I\right]  & AB^{\ast}=BA^{\ast}\\
\left[  II\right]  & \text{the }n\times2n\text{ matrix }\left(  AB\right)
\text{ has maximal rank}%
\end{array}
\]
Denote with $W$ the set of all couples of complex $n\times n$ matrices
fulfilling $\left[  I\right]  $ and $\left[  II\right]  $
\begin{equation}
W=\left\{  \left(  A,B\right)  \left\vert \,\left[  I\right]  ,\,\left[
II\right]  \right.  \right\}  \label{W}%
\end{equation}
The following Proposition is an immediate consequence of Theorem 3.1.4 in
\cite{Gorba}

\begin{proposition}
\label{Panka}Let $H_{0}$ be a closed symmetric operator with equal deficiency
indices $\left(  n,n\right)  $, $n<\infty$, acting on a Hilbert space, and let
$\left(  \mathbb{C}^{n},\Gamma_{1},\Gamma_{2}\right)  $ be its boundary value
space. There is a bijective correspondence between the selfadjoint extensions
of $H_{0}$ and the set $W$. A selfadjoint extension $H^{A,B}$, corresponding
to $\left(  A,B\right)  \in W$, is given by the restriction of $H_{0}^{\ast}$
to those elements $\psi\in D(H_{0}^{\ast})$ satisfying the boundary conditions%
\begin{equation}
A\Gamma_{1}\psi=B\Gamma_{2}\psi\label{s.a.b.c.}%
\end{equation}

\end{proposition}

In what follows, we will exploit this result to construct selfadjoint
operators associated to point perturbations of the Laplacian in bounded
domains. We consider the symmetric operator%
\begin{equation}
\left\{
\begin{array}
[c]{l}%
\bigskip D\left(  H_{0}\right)  =\left\{  \psi\in H^{2}\cap H_{0}^{1}%
(\Omega)\left\vert \ \left.  \psi\right\vert _{\left\{  \underline{x}%
_{k}\right\}  _{k=1}^{N}}=0\right.  \right\} \\
H_{0}\psi=-\Delta\psi
\end{array}
\right.  \label{H_0}%
\end{equation}
where $\Omega$ is an open bounded domain of $\mathbb{R}^{3}$ with regular
boundary (for instance $\partial\Omega$ of class $C^{1}$ and bounded) and
$\left\{  \underline{x}_{k}\right\}  _{k=1}^{N}$ is a finite set of points in
$\Omega$. The defect spaces of $H_{0}$, have the following characterization

\begin{lemma}
\label{Defect}Let $\mathcal{H}_{\pm}=Ker\left(  \pm i-H_{0}^{\ast}\right)  $
be the defect spaces related to the operator $H_{0}$ defined in (\ref{H_0});
then%
\[
\mathcal{H}_{+}=l.s.\left\{  \mathcal{G}_{0}^{-i,k},\ k=1...N\right\}
\]%
\[
\mathcal{H}_{-}=l.s.\left\{  \mathcal{G}_{0}^{i,k},\ k=1...N\right\}
\]
where $\mathcal{G}_{0}^{\pm i,k}$ are the integral kernels of $\left(
-\Delta\pm i\right)  $%
\begin{equation}
\left(  -\Delta\pm i\right)  \mathcal{G}_{0}^{\pm i,k}=\delta\left(
\cdot-\underline{x}_{k}\right)  \qquad k=1...N \label{G_0 def}%
\end{equation}
with Dirichlet conditions on the boundary of $\Omega$, while $l.s.$ denotes
the linear span of the sets.
\end{lemma}

\begin{proof}
The spaces $\mathcal{H}_{+}$ and $\mathcal{H}_{-}$ are described by the
following conditions
\begin{equation}
\psi\in\mathcal{H}_{+}\Rightarrow\left\{
\begin{array}
[c]{l}%
\left(  \psi,\,\left(  \Delta+i\right)  \varphi\right)  _{L^{2}(\Omega)}=0\\
\psi\in L^{2}(\Omega)
\end{array}
\right.  \qquad\forall\varphi\in D(H_{0}) \label{defect+}%
\end{equation}
and%
\begin{equation}
\psi\in\mathcal{H}_{-}\Rightarrow\left\{
\begin{array}
[c]{l}%
\left(  \psi,\,\left(  \Delta-i\right)  \varphi\right)  _{L^{2}(\Omega)}=0\\
\psi\in L^{2}(\Omega)
\end{array}
\right.  \qquad\forall\varphi\in D(H_{0}) \label{defect-}%
\end{equation}

We analyze the first of these problems. Consider a collection of open balls
$B_{k}$ centered in the points $\left\{  \underline{x}_{k}\right\}  _{k=1}%
^{N}$ with $\cup_{k=1}^{N}B_{k}\subset\Omega$ and $B_{k}\cap B_{k^{\prime}%
}=\varnothing$ $\forall k\neq k^{\prime}$. Given any function $\varphi\in
H^{2}\cap H_{0}^{1}(\Omega)$, we define the set $U_{\varphi}$
\begin{equation}
\xi\in U_{\varphi}\rightarrow\xi(\underline{x})=\left\{
\begin{array}
[c]{l}%
0\qquad\underline{x}\in\Omega\backslash\cup_{k=1}^{N}B_{k}\\
\varphi_{k}(\underline{x}),\ \varphi_{k}\in\left\{  H^{2}\cap H_{0}^{1}%
(B_{k})\,\left\vert \varphi_{k}(\underline{x}_{k})=\varphi(\underline{x}%
_{k})\right.  \right\}  \quad otherwise
\end{array}
\right.  \label{csi}%
\end{equation}
Given any $\varphi^{0}\in D(H_{0})$ there are infinitely many $\varphi\in$
$H^{2}\cap H_{0}^{1}(\Omega)$ coinciding with $\varphi^{0}$ in $\Omega
\backslash\cup_{k=1}^{N}B_{k}$; for each one of them the difference
$\xi_{\varphi}=\varphi-\varphi^{0}$ belongs to $U_{\varphi}$. Then the whole
domain $D(H_{0})$ can be represented in the form%
\begin{equation}
\varphi^{0}\in D(H_{0})\rightarrow\varphi^{0}=\varphi-\xi_{\varphi}
\label{D(H_0)}%
\end{equation}
by varying $\varphi$ in $H^{2}\cap H_{0}^{1}(\Omega)$ and $\xi$ in
$U_{\varphi}$. Making use of this representation, equation (\ref{defect+}) can
be written in the form%
\begin{equation}
\left\{
\begin{array}
[c]{l}%
\left(  \psi,\,\left(  \Delta+i\right)  \varphi\right)  _{L^{2}(\Omega
)}-\left(  \psi,\,\left(  \Delta+i\right)  \xi_{\varphi}\right)
_{L^{2}(\Omega)}=0\\
\psi\in L^{2}(\Omega)
\end{array}
\right.  \qquad\forall\left(  \varphi,\xi\right)  \in H^{2}\cap H_{0}%
^{1}(\Omega)\times U_{\varphi} \label{defect+ 1}%
\end{equation}
Due to the essential selfadjointness of the operator $-\Delta$ in the space
$H^{2}\cap H_{0}^{1}(\Omega)$, we know that the equation%
\[
\left(  \psi,\,\left(  \Delta+i\right)  \varphi\right)  _{L^{2}(\Omega
)}=0\qquad\forall\varphi\in H^{2}\cap H_{0}^{1}(\Omega)
\]
has no solution in $L^{2}(\Omega)$; then there is no $\psi$ for which both
terms of (\ref{defect+ 1}) are zero. On the other hand, being $\varphi$ and
$\xi$ linked only by their values in the points $\left\{  \underline{x}%
_{k}\right\}  _{k=1}^{N}$, the only non trivial solution of (\ref{defect+}) is
a linear combination of the Green's functions $\mathcal{G}_{0}^{-i,k}$ and of
their derivatives. Nevertheless we notice that, in the case of 3-D domains,
none of the functions $\nabla^{n}\mathcal{G}_{0}^{-i,k}$ belongs to
$L^{2}(\Omega)$. The observations above implies that $\mathcal{H}_{+}$ is an
$N$ dimensional space generated by the functions $\left\{  \mathcal{G}%
_{0}^{-i,k},\ k=1...N\right\}  $. Proceeding in the same way, it is easy to
obtain a similar characterization for $\mathcal{H}_{-}$, that is:
$\mathcal{H}_{-}=l.s.\left\{  \mathcal{G}_{0}^{i,k},\ k=1...N\right\}  $.
\end{proof}

In the following $\mathcal{G}_{0}^{z,k}$ shall denote the integral kernel of
$\left(  -\Delta+z\right)  $ with Dirichlet conditions on the boundary of
$\Omega$ and centered in the points $\left\{  \underline{x}_{k}\right\}
_{k=1}^{N}$; it is worthwhile to recall that these functions are properly
defined by\footnote{Here and in the following we use $\operatorname{Re}%
\sqrt{z}\geq0$ as determination of the square root in the complex plane.}%
\begin{equation}
\mathcal{G}_{0}^{z,k}(\underline{x})=\frac{e^{-\sqrt{z}\left\vert
\underline{x}-\underline{x}_{k}\right\vert }}{4\pi\left\vert \underline
{x}-\underline{x}_{k}\right\vert }-h_{z,k}(\underline{x}) \label{G_0(z) def}%
\end{equation}%
\begin{equation}
\left\{
\begin{array}
[c]{l}%
\medskip\left(  -\Delta+z\right)  h_{z,k}=0\\
\left.  h_{z,k}\right\vert _{\partial\Omega}=\left.  \frac{e^{-\sqrt
{z}\left\vert \underline{x}-\underline{x}_{k}\right\vert }}{4\pi\left\vert
\underline{x}-\underline{x}_{k}\right\vert }\right\vert _{\partial\Omega}%
\end{array}
\right.  \label{h_(z,k)}%
\end{equation}
whenever $-z$ does not belongs to the point spectrum of the Dirichlet
Laplacian in $\Omega$. In Section 3, the asymptotic properties of $h_{z,k}$,
as $z$ approaches spectral points, are considered.

Our next task is to defines a couple of operators $\Gamma_{i=1,2}%
:D(H_{0}^{\ast})\rightarrow\mathbb{C}^{N}$ such that the triple $\left(
\mathbb{C}^{N},\Gamma_{1},\Gamma_{2}\right)  $ forms a boundary value space
for $H_{0}$. To this aim, and following the analogous definitions given for
the case $\Omega=\mathbb{R}^{3}$ (see for instance \cite{Panka}-
\cite{Kurasov}, and \cite{Caccia 1}), we define%
\begin{equation}
\left(  \Gamma_{1}\psi\right)  _{j}=\lim_{\underline{x}\rightarrow
\underline{x}_{j}}4\pi\left\vert \underline{x}-\underline{x}_{j}\right\vert
\psi(\underline{x})\qquad j=1...N \label{Gamma_1}%
\end{equation}%
\begin{equation}
\left(  \Gamma_{2}\psi\right)  _{j}=\lim_{\underline{x}\rightarrow
\underline{x}_{j}}\left(  \psi(\underline{x})-\frac{\left(  \Gamma_{1}%
\psi\right)  _{j}}{4\pi\left\vert \underline{x}-\underline{x}_{j}\right\vert
}\right)  \qquad j=1...N \label{Gamma_2}%
\end{equation}

\begin{theorem}
\label{BVT}The triple $\left(  \mathbb{C}^{N},\Gamma_{1},\Gamma_{2}\right)  $
defined by (\ref{Gamma_1}) and (\ref{Gamma_2}) forms a boundary value space
for $H_{0}$.
\end{theorem}

\begin{proof}
The Von Neumann decomposition formula of the domain $D(H_{0}^{\ast})$ (see for
instance \cite{Simon}) allow us to write the generic vector $\psi\in
D(H_{0}^{\ast})$ in the form%
\begin{equation}
\psi=\psi_{0}+\sum_{k=1}^{N}\left(  a_{k}\mathcal{G}_{0}^{-i,k}+b_{k}%
\mathcal{G}_{0}^{i,k}\right)  \qquad a_{k},b_{k}\in\mathbb{C};\ \psi_{0}\in
D(H_{0}) \label{V.Neumann_rep}%
\end{equation}
We will use this representation in order to prove that relation
(\ref{b.v.s. 1}) holds under our assumptions. Let $\psi,\phi\in D(H_{0}^{\ast
})$ be given by%
\begin{equation}
\psi=\psi_{0}+\sum_{k=1}^{N}\left(  a_{k}\mathcal{G}_{0}^{-i,k}+b_{k}%
\mathcal{G}_{0}^{i,k}\right)  ;\quad\phi=\phi_{0}+\sum_{k=1}^{N}\left(
\alpha_{k}\mathcal{G}_{0}^{-i,k}+\beta_{k}\mathcal{G}_{0}^{i,k}\right)
\label{vettori}%
\end{equation}
From the symmetry of $H_{0}$ it is easy to obtain%
\[
\left(  \psi,H_{0}^{\ast}\varphi\right)  -\left(  H_{0}^{\ast}\psi
,\varphi\right)  =-2i\sum_{k,j=1}^{N}\left(  a_{k}\alpha_{j}^{\ast}-b_{j}%
\beta_{k}^{\ast}\right)  \left(  \mathcal{G}_{0}^{-i,k},\mathcal{G}_{0}%
^{-i,j}\right)
\]
Making use of the properties of the functions $\left\{  \mathcal{G}_{0}%
^{-i,k}\right\}  _{k=1}^{N}$, we have%
\[
\left(  \mathcal{G}_{0}^{-i,k},\mathcal{G}_{0}^{-i,j}\right)  =\left\{
\begin{array}
[c]{l}%
-\frac{i}{2}\left(  \mathcal{G}_{0}^{-i,k}(\underline{x}_{j})-\mathcal{G}%
_{0}^{i,j}\left(  \underline{x}_{k}\right)  \right)  \qquad k\neq j\\
-\operatorname{Re}ih_{i,k}(\underline{x}_{k})+\frac{i}{2}\frac{1}{4\pi}\left(
\sqrt{-i}-\sqrt{i}\right)  \qquad k=j
\end{array}
\right.
\]
Taking into account that $h_{-i,k}=\left(  h_{i,k}\right)  ^{\ast}$, the
previous relation can be written as follows%
\begin{gather}
\left(  \psi,H_{0}^{\ast}\varphi\right)  -\left(  H_{0}^{\ast}\psi
,\varphi\right)  =-\sum_{k\neq j}\left(  a_{k}\alpha_{j}^{\ast}-b_{k}\beta
_{j}^{\ast}\right)  \left(  \mathcal{G}_{0}^{-i,k}(\underline{x}%
_{j})-\mathcal{G}_{0}^{i,j}\left(  \underline{x}_{k}\right)  \right)
+\nonumber\\
+\frac{1}{4\pi}\sum_{j}\left(  a_{j}\alpha_{j}^{\ast}-b_{j}\beta_{j}^{\ast
}\right)  \left(  \sqrt{-i}-\sqrt{i}\right)  +\sum_{j}\left(  a_{j}\alpha
_{j}^{\ast}-b_{j}\beta_{j}^{\ast}\right)  \left(  h_{-i,j}(\underline{x}%
_{j})-h_{i,j}(\underline{x}_{j})\right)  \label{rel1}%
\end{gather}
On the other hand, making use of the representation (\ref{V.Neumann_rep}), the
action of the operators (\ref{Gamma_1}) and (\ref{Gamma_2}) is explicitly
given by%
\begin{equation}
\left(  \Gamma_{1}\psi\right)  _{j}=a_{j}+b_{j} \label{Gamma_1(1)}%
\end{equation}%
\begin{equation}
\left(  \Gamma_{2}\psi\right)  _{j}=\sum_{\substack{k\\k\neq j}}a_{k}%
\mathcal{G}_{0}^{-i,k}(\underline{x}_{j})+b_{k}\mathcal{G}_{0}^{i,k}%
(\underline{x}_{j})-\frac{1}{4\pi}\left[  a_{j}\sqrt{-i}+b_{j}\sqrt{i}+\left(
\Gamma_{1}\psi\right)  _{j}\right]  -a_{k}h_{-i,j}(\underline{x}_{j}%
)-b_{k}h_{i,j}(\underline{x}_{j}) \label{Gamma_2(1)}%
\end{equation}
and the second member of (\ref{b.v.s. 1}) reads as%
\begin{gather*}
\left\langle \Gamma_{1}\psi,\Gamma_{2}\varphi\right\rangle -\left\langle
\Gamma_{2}\psi,\Gamma_{1}\varphi\right\rangle =-\sum_{k\neq j}\left(
a_{k}\alpha_{j}^{\ast}-b_{k}\beta_{j}^{\ast}\right)  \left(  \mathcal{G}%
_{0}^{-i,k}(\underline{x}_{j})-\mathcal{G}_{0}^{i,j}\left(  \underline{x}%
_{k}\right)  \right)  +\\
-\sum_{k\neq j}a_{k}\beta_{j}^{\ast}\left(  \mathcal{G}_{0}^{-i,k}%
(\underline{x}_{j})-\mathcal{G}_{0}^{-i,j}\left(  \underline{x}_{k}\right)
\right)  -\sum_{k\neq j}b_{k}\alpha_{j}^{\ast}\left(  \mathcal{G}_{0}%
^{i,k}(\underline{x}_{j})-\mathcal{G}_{0}^{i,j}\left(  \underline{x}%
_{k}\right)  \right)  +\\
+\frac{1}{4\pi}\sum_{j}\left(  a_{j}\alpha_{j}^{\ast}-b_{j}\beta_{j}^{\ast
}\right)  \left(  \sqrt{-i}-\sqrt{i}\right)  +\sum_{j}\left(  a_{j}\alpha
_{j}^{\ast}-b_{j}\beta_{j}^{\ast}\right)  \left(  h_{-i,j}(\underline{x}%
_{j})-h_{i,j}(\underline{x}_{j})\right)
\end{gather*}
Concerning the difference $\mathcal{G}_{0}^{\pm i,k}(\underline{x}%
_{j})-\mathcal{G}_{0}^{\pm i,j}\left(  \underline{x}_{k}\right)  $, from the
explicit expression of $\mathcal{G}_{0}^{\pm i,k}$, follows
\[
\mathcal{G}_{0}^{\pm i,k}(\underline{x}_{j})-\mathcal{G}_{0}^{\pm i,j}\left(
\underline{x}_{k}\right)  =h_{\pm i,k}(\underline{x}_{j})-h_{\pm
i,j}(\underline{x}_{k})
\]
and a simple calculation, exploiting the definition (\ref{G_0 def}), shows
that%
\[
\left(  \mathcal{G}_{0}^{\pm i,k},\mathcal{G}_{0}^{\mp i,j}\right)  =\left(
\mathcal{G}_{0}^{\pm i,k},\mathcal{G}_{0}^{\mp i,j}\right)  +i\left(  h_{\pm
i,k}(\underline{x}_{j})-h_{\pm i,j}(\underline{x}_{k})\right)  \rightarrow
\left(  h_{\pm i,k}(\underline{x}_{j})-h_{\pm i,j}(\underline{x}_{k})\right)
=0
\]
from which we recover%
\[
\left(  \psi,H_{0}^{\ast}\varphi\right)  -\left(  H_{0}^{\ast}\psi
,\varphi\right)  =\left\langle \Gamma_{1}\psi,\Gamma_{2}\varphi\right\rangle
-\left\langle \Gamma_{2}\psi,\Gamma_{1}\varphi\right\rangle
\]

The proof of (\ref{b.v.s. 2}) easily follows from definitions
(\ref{Gamma_1(1)}) and (\ref{Gamma_2(1)}).
\end{proof}

The above result allows a general characterization of the selfadjoint
extensions of operator $H_{0}$ in terms of selfadjoint boundary conditions.
From Proposition \ref{Panka}, any selfadjoint extension of $H_{0}$ can be
parametrized through an element of the set $W$ (see definition (\ref{W})); the
one corresponding to the couple $\left(  A,B\right)  \in W$ is given by the
restriction of $H_{0}^{\ast}$ to those elements $\psi\in D(H_{0}^{\ast})$
which satisfy the boundary conditions (\ref{s.a.b.c.}). Denoting with $H^{AB}$
this extension, we have%
\begin{equation}
D(H^{AB})=\left\{  \psi\in D(H_{0}^{\ast})\left\vert \,A\Gamma_{1}\psi
=B\Gamma_{2}\psi\right.  \right\}  \label{H_AB_dominio(0)}%
\end{equation}%
\begin{equation}
H^{AB}\psi=H_{0}^{\ast}\psi\label{H_AB_def}%
\end{equation}

The next theorem gives an explicit representation of the domain and provide a
resolvent formula for any operator of type (\ref{H_AB_dominio(0)}%
)-(\ref{H_AB_def}).

\begin{theorem}
\label{Theorem H_AB}Fix $\left(  A,B\right)  \in W$, let $H^{AB}$ - defined by
(\ref{H_AB_dominio(0)})-(\ref{H_AB_def}) - be the related extension and
$R_{z}^{AB}$ its resolvent. For any $\lambda\in\mathbb{C}\backslash\mathbb{R}%
$, the following representation holds%
\begin{align}
D(H^{AB})  &  =\left\{  \psi\in L^{2}(\Omega)\right\vert \,\psi=\phi^{\lambda
}+%
{\textstyle\sum\nolimits_{k=1}^{N}}
q_{k}\mathcal{G}_{0}^{\lambda,k},\ \phi^{\lambda}\in H^{2}\cap H_{0}%
^{1}(\Omega),\nonumber\\
&  \qquad q_{j}=\left(  \Gamma_{1}\psi\right)  _{j},\ \sum_{j}B_{kj}%
\phi^{\lambda}(\underline{x}_{j})=\sum_{j}\left(  B\,\Gamma(\lambda)+A\right)
_{kj}\left.  q_{j}\right\}  \label{H_AB_dom}%
\end{align}%
\begin{equation}
H^{AB}\psi=-\Delta\phi-\lambda%
{\textstyle\sum\nolimits_{k=1}^{N}}
q_{k}\mathcal{G}_{0}^{\lambda,k} \label{H_AB}%
\end{equation}%
\begin{equation}
R_{z}^{AB}\varphi=R_{z}\varphi+\sum_{j,l,k=1}^{N}\left(  B\Gamma(z)+A\right)
_{jl}^{-1}B_{lk}\,R_{z}\varphi(\underline{x}_{k})\,\mathcal{G}_{0}^{z,j}%
,\quad\forall\varphi\in L^{2}(\Omega),\ z\in\mathbb{C}\backslash\mathbb{R}
\label{R_AB}%
\end{equation}%
\begin{equation}
\Gamma_{kj}(z)=\left\{
\begin{array}
[c]{l}%
-\mathcal{G}_{0}^{z,j}(\underline{x}_{k})\qquad j\neq k\\
h_{z,k}(\underline{x}_{k})+\frac{\sqrt{z}}{4\pi}\qquad j=k
\end{array}
\right.  \label{Gamma_kj}%
\end{equation}
where $R_{z}=\frac{1}{-\Delta+z}$ is the resolvent operator associated to
$-\Delta$ with Dirichlet boundary conditions in $\Omega$.
\end{theorem}

\begin{proof}
As follows from the decomposition formula (\ref{V.Neumann_rep}), the generic
vector in the domain $D(H_{0}^{\ast})$ has the form $\psi=\psi_{0}+\sum
_{k=1}^{N}\left(  a_{k}\mathcal{G}_{0}^{-i,k}+b_{k}\mathcal{G}_{0}%
^{i,k}\right)  $ with $\psi_{0}\in D(H_{0})$. Fix $\lambda\in\mathbb{C}%
\backslash\mathbb{R}$ and set $\phi^{\lambda}$%
\begin{equation}
\phi^{\lambda}=\psi_{0}+\sum_{k=1}^{N}\left(  a_{k}\mathcal{G}_{0}%
^{-i,k}+b_{k}\mathcal{G}_{0}^{i,k}\right)  -\sum_{k=1}^{N}q_{k}\mathcal{G}%
_{0}^{\lambda,k} \label{Phi_lambda 1}%
\end{equation}%
\begin{equation}
q_{j}=\left(  \Gamma_{1}\psi\right)  _{j}=a_{j}+b_{j} \label{BC 1}%
\end{equation}
Our first task is to show that $\phi^{\lambda}\in H^{2}\cap H_{0}^{1}(\Omega
)$; to this aim we notice that $\phi^{\lambda}\in H^{2}\cap H_{0}^{1}%
(\Omega\backslash\left\{  \underline{x}_{k}\right\}  _{k=1}^{N})$, so that the
set of its singular points is contained in $\left\{  \underline{x}%
_{k}\right\}  _{k=1}^{N}$. Taking into account the explicit expressions of the
Green functions $\mathcal{G}_{0}^{z,k}$, it is easy to see that, apart from
regular terms, $\phi^{\lambda}$ behaves around $\underline{x}_{\bar{k}}%
\in\left\{  \underline{x}_{k}\right\}  _{k=1}^{N}$ as%
\[
\phi_{s,\bar{k}}(\underline{x})=a_{\bar{k}}\frac{e^{-\sqrt{-i}\left\vert
\underline{x}-\underline{x}_{\bar{k}}\right\vert }}{4\pi\left\vert
\underline{x}-\underline{x}_{\bar{k}}\right\vert }+b_{\bar{k}}\frac
{e^{-\sqrt{i}\left\vert \underline{x}-\underline{x}_{\bar{k}}\right\vert }%
}{4\pi\left\vert \underline{x}-\underline{x}_{\bar{k}}\right\vert }-q_{\bar
{k}}\frac{e^{-\sqrt{\lambda}\left\vert \underline{x}-\underline{x}_{\bar{k}%
}\right\vert }}{4\pi\left\vert \underline{x}-\underline{x}_{\bar{k}%
}\right\vert }%
\]
A direct calculation shows that $\phi_{s,\bar{k}}$ and $\nabla\phi_{s,\bar{k}%
}$ both have finite values in the limit $\underline{x}\rightarrow\underline
{x}_{\bar{k}}$%
\begin{equation}
\lim_{\underline{x}\rightarrow\underline{x}_{\bar{k}}}\phi_{s,\bar{k}}%
=-\frac{1}{4\pi}\left[  a_{\bar{k}}\sqrt{-i}+b_{\bar{k}}\sqrt{i}-q_{\bar{k}%
}\sqrt{\lambda}\right]  \label{Phi_lambda 2}%
\end{equation}%
\begin{equation}
\lim_{\underline{x}\rightarrow\underline{x}_{\bar{k}}}\nabla\phi_{s,\bar{k}%
}=\hat{r}_{\bar{k}}\,\frac{1}{8\pi}\left[  a_{\bar{k}}\sqrt{-i}+b_{\bar{k}%
}\sqrt{i}-q_{\bar{k}}\lambda\right]
\end{equation}
with $\hat{r}_{\bar{k}}=\frac{\underline{x}_{\bar{k}}}{\left\vert
\underline{x}_{\bar{k}}\right\vert }$. Moreover, a compensation mechanism
avoid the function $\Delta\phi_{s,\bar{k}}$ to have purely distributional
terms. Infact, making use of the equation%
\[
\left(  -\Delta+z\right)  \frac{e^{-\sqrt{z}\left\vert \underline
{x}-\underline{x}_{k}\right\vert }}{4\pi\left\vert \underline{x}-\underline
{x}_{k}\right\vert }=\delta\left(  \underline{x}-\underline{x}_{k}\right)
\]
holding for all $z\in\mathbb{C}\backslash\left\{  \lambda_{n}\right\}
_{n\in\mathbb{N}}$, we have%
\begin{align*}
\Delta\phi_{s,\bar{k}}  &  =\Delta\left[  a_{\bar{k}}\frac{e^{-\sqrt
{-i}\left\vert \underline{x}-\underline{x}_{\bar{k}}\right\vert }}%
{4\pi\left\vert \underline{x}-\underline{x}_{\bar{k}}\right\vert }+b_{\bar{k}%
}\frac{e^{-\sqrt{i}\left\vert \underline{x}-\underline{x}_{\bar{k}}\right\vert
}}{4\pi\left\vert \underline{x}-\underline{x}_{\bar{k}}\right\vert }%
-q_{\bar{k}}\frac{e^{-\sqrt{\lambda}\left\vert \underline{x}-\underline
{x}_{\bar{k}}\right\vert }}{4\pi\left\vert \underline{x}-\underline{x}%
_{\bar{k}}\right\vert }\right]  =\\
&  =-a_{\bar{k}}\,i\,\frac{e^{-\sqrt{-i}\left\vert \underline{x}-\underline
{x}_{\bar{k}}\right\vert }}{4\pi\left\vert \underline{x}-\underline{x}%
_{\bar{k}}\right\vert }+b_{\bar{k}}\,i\,\frac{e^{-\sqrt{i}\left\vert
\underline{x}-\underline{x}_{\bar{k}}\right\vert }}{4\pi\left\vert
\underline{x}-\underline{x}_{\bar{k}}\right\vert }-q_{\bar{k}}\,\lambda
\,\frac{e^{-\sqrt{\lambda}\left\vert \underline{x}-\underline{x}_{\bar{k}%
}\right\vert }}{4\pi\left\vert \underline{x}-\underline{x}_{\bar{k}%
}\right\vert }+\left(  -a_{\bar{k}}-b_{\bar{k}}+q_{\bar{k}}\right)
\delta(\underline{x}-\underline{x}_{k})
\end{align*}
From (\ref{BC 1}) it follows that $\Delta\phi_{s,\bar{k}}\in L^{2}(\Omega)$.
The previous observations imply that $\phi^{\lambda}\in H^{2}\cap H_{0}%
^{1}(S)$; this result guarantees the equivalence of the representations
\begin{equation}
\psi=\psi_{0}+\sum_{k=1}^{N}\left(  a_{k}\mathcal{G}_{0}^{-i,k}+b_{k}%
\mathcal{G}_{0}^{i,k}\right)  ,\quad\psi_{0}\in D(H_{0}) \label{Von Neumann}%
\end{equation}
and%
\begin{equation}
\psi=\phi^{\lambda}+%
{\displaystyle\sum_{k=1}^{N}}
q_{k}\mathcal{G}_{0}^{\lambda,k},\quad\lambda\in\mathbb{C}\backslash
\mathbb{R}^{+},\ \phi^{\lambda}\in H^{2}\cap H_{0}^{1}(S)
\label{Von Neumann_alter}%
\end{equation}
of the domain $D(H_{0}^{\ast})$.

From (\ref{Phi_lambda 2}) and making use of the explicit expression
(\ref{G_0(z) def})-(\ref{Phi_lambda 1}), it follows that any $\psi$ in
(\ref{H_AB_dom}) obeys to the relation $\phi^{\lambda}(\underline{x}%
_{k})=\Gamma_{kj}(\lambda)q_{j}+\left(  \Gamma_{2}\psi\right)  _{k}$, with
$\Gamma_{kj}(\lambda)$ given by (\ref{Gamma_kj}) and $q_{j}=\left(  \Gamma
_{1}\psi\right)  _{j}$; taking into account the boundary relations
$A\Gamma_{1}\psi=B\Gamma_{2}\psi$, we get the following characterization of
the values of $\phi^{\lambda}$ in the points $\underline{x}_{k}$
\begin{equation}
\sum_{j}B_{kj}\phi^{\lambda}(\underline{x}_{j})=\sum_{j}\left(  B\,\Gamma
(\lambda)+A\right)  _{kj}q_{j} \label{BC 2}%
\end{equation}
Relations (\ref{Von Neumann_alter})-(\ref{BC 2}) identify the alternative
representation of the domain $D(H^{AB})$ given in (\ref{H_AB_dom}).

Let $\psi=\phi^{\lambda}+%
{\textstyle\sum\nolimits_{k=1}^{N}}
q_{k}\mathcal{G}_{0}^{\lambda,k}$: from the definition (\ref{H_AB_def}), we
get for the action of operator $H^{AB}$%
\[
\left(  H^{AB}\psi,\varphi\right)  =\left(  \phi^{\lambda},H_{0}%
\varphi\right)  +\left(
{\textstyle\sum\nolimits_{k=1}^{N}}
q_{k}\mathcal{G}_{0}^{\lambda,k},H_{0}\varphi\right)
\]
then an integration by parts easily shows that $H^{AB}\psi$ is described by
the relation (\ref{H_AB}).

The resolvent $R_{z}^{AB}$ associated to the operator $H^{AB}$, for
$z\in\mathbb{C}\backslash\mathbb{R}$, is a bounded linear map: $L^{2}%
(\Omega)\rightarrow D(H^{AB})$. Let $\varphi\in L^{2}(\Omega)$, exploiting the
regularity of the map $R_{z}\in\mathcal{B}(L^{2}(\Omega),H^{2}\cap H_{0}%
^{1}(\Omega))$, we can identify with $R_{z}\varphi$ the $H^{2}$-part of the
function $R_{z}^{AB}\varphi$; therefore, from (\ref{Von Neumann_alter}%
)-(\ref{BC 2}), the function $R_{z}^{AB}\varphi\in D(H^{AB})$ has the
following representation%
\[
R_{z}^{AB}\varphi=R_{z}\varphi+%
{\textstyle\sum\nolimits_{k=1}^{N}}
q_{k}(\varphi)\mathcal{G}_{0}^{z,k}%
\]%
\[
\sum_{j}B_{kj}R_{z}\varphi(\underline{x}_{j})=\sum_{l,j}B_{kl}\Gamma
_{lj}(z)q_{j}(\varphi)+\sum_{j}A_{kj}q_{j}(\varphi)
\]
from which it follows%
\begin{equation}
q_{j}(\varphi)=\sum_{l,k}\left(  B\Gamma(z)+A\right)  _{jl}^{-1}B_{lk}%
\,R_{z}\varphi(\underline{x}_{k}) \label{Risolvente2}%
\end{equation}

\end{proof}

\section{Point interactions operators in bounded domains}

The family of extensions described by (\ref{H_AB_dom})-(\ref{H_AB}) include
both the, so called, local and non local point interactions. The interaction
type is specified by the boundary conditions (\ref{s.a.b.c.}); from the
definitions of $\Gamma_{1}$ and $\Gamma_{2}$ it follows that the asymptotics
of $\psi\in D\left(  H_{0}^{\ast}\right)  $ as $\underline{x}\rightarrow
\underline{x}_{k}$ is given by%
\[
\psi\underset{\underline{x}\rightarrow\underline{x}_{k}}{\sim}\frac{1}%
{4\pi\left\vert \underline{x}-\underline{x}_{k}\right\vert }\left(  \Gamma
_{1}\psi\right)  _{k}+\left(  \Gamma_{2}\psi\right)  _{k}+o(1)
\]
then, through the relation $A\Gamma_{1}\psi=B\Gamma_{2}\psi$, a link between
the asymptotic behavior of the singular and the regular part of the state
$\psi$ is established. Whenever the coefficient of the singularity in the
point $\underline{x}_{k}$, $\left(  \Gamma_{1}\psi\right)  _{k}$, depends only
on the value in $\underline{x}_{k}$ of the regular part of $\psi$, i.e on
$\left(  \Gamma_{2}\psi\right)  _{k}$, the interaction will be said to be
local. This condition is obtained by requiring the matrices $A$ and $B$ to be
diagonal. Without loss of generality, we can parametrize the most general
extension of local kind using a set of $N$ real, $\left\{  \alpha_{k}\right\}
_{k=1}^{N}$, such that%
\begin{equation}
A_{ij}=\alpha_{i}\,\delta_{ij},\quad B_{ij}=\delta_{ij} \label{local}%
\end{equation}
Under this assumption, we shall investigate some spectral properties of
operators $H^{AB}$. Taking in account (\ref{local}), the resolvent formula
(\ref{R_AB}) can be written as%
\begin{equation}
R_{z}^{AB}\varphi=R_{z}\varphi+\sum_{j,k=1}^{N}\left(  \Gamma(z)+A\right)
_{jk}^{-1}\,R_{z}\varphi(\underline{x}_{k})\,\mathcal{G}_{0}^{z,j}
\label{R_AB 1}%
\end{equation}
with%
\begin{equation}
\left(  \Gamma(z)+A\right)  _{kj}=\left\{
\begin{array}
[c]{l}%
-\mathcal{G}_{0}^{z,j}(\underline{x}_{k})\qquad j\neq k\\
\alpha_{k}+h_{z,k}(\underline{x}_{k})+\frac{\sqrt{z}}{4\pi}\qquad j=k
\end{array}
\right.  \label{Gamma_kj 1}%
\end{equation}

In what follows, we shall make use of the spectral properties of the Laplacian
operator defined in $H^{2}\cap H_{0}^{1}(\Omega)$. Let us denote with
$\left\{  \lambda_{n}\right\}  _{n\in\mathbb{N}}$ the spectral points of
$-\Delta$ with Dirichlet boundary conditions in $\Omega$. We shall assume each
eigenvalue $\lambda_{n}$ to have a degeneracy $M_{n}$; the normalized
eigenvectors related to $\lambda_{n}$, denoted with $\left\{  \phi
_{n,m}\right\}  _{\mathbb{\,}m\,=1}^{M_{n}}$, satisfy the equations
\begin{equation}
\left\{
\begin{array}
[c]{l}%
\left(  H-\lambda_{n}\right)  \phi_{n,m}=0\\
\left.  \phi_{n,m}\right\vert _{\partial\Omega}=0;\qquad\left\Vert \phi
_{n,m}\right\Vert _{L^{2}(\Omega)}=1
\end{array}
\right.  \label{stati stazionari}%
\end{equation}
These vectors span an $M_{n}$ dimensional space in $H^{2}\cap H_{0}^{1}%
(\Omega)$.

The resolvent operator $R_{z}=\frac{1}{-\Delta+z}$ is a bounded map:
$L^{2}(\Omega)\rightarrow H^{2}\cap H_{0}^{1}(\Omega)$ whose action on a
vector $\varphi$ can be represented as follows%
\begin{equation}
R_{z}\varphi=\sum_{n\in\mathbb{N}}\sum_{m=1}^{M_{n}}\frac{\left(  \varphi
,\phi_{n,m}\right)  }{\lambda_{n}+z}\phi_{n,m} \label{Res free}%
\end{equation}

The next Lemma gives a general characterization of the asymptotic behavior of
the functions $h_{z,k}$ as $z$ approximates a point in the spectrum of
operator $-\Delta$.

\begin{lemma}
\label{Lemma!}Let $\lambda_{\bar{n}}$ be an eigenvalue of the operator
$-\Delta$ - defined on $H^{2}\cap H_{0}^{1}(\Omega)$ - and $\phi_{\bar{n},m}$
be the corresponding eigenfunctions and $h_{z,k}$ defined by (\ref{h_(z,k)});
then the following limit holds%
\begin{equation}
\lim_{\left\vert \varepsilon\right\vert \rightarrow0}\left\Vert \varepsilon
h_{-\lambda_{\bar{n}}+\varepsilon,\,k}+\sum_{m=1}^{M_{\bar{n}}}\phi_{\bar
{n},m}^{\ast}(\underline{x}_{k})\,\phi_{\bar{n},m}\right\Vert _{L^{2}(\Omega
)}=0 \label{h_(z,k)_lim}%
\end{equation}
with $\varepsilon\in\mathbb{C}$.
\end{lemma}

\begin{proof}
From definition (\ref{G_0(z) def}), we get%
\[
\left\Vert \varepsilon h_{-\lambda_{\bar{n}}+\varepsilon,\,k}+\sum
_{m=1}^{M_{\bar{n}}}\phi_{\bar{n},m}^{\ast}(\underline{x}_{k})\,\phi_{\bar
{n},m}\right\Vert _{L^{2}(\Omega)}\leq\left\vert \varepsilon\right\vert
\left\Vert \frac{e^{-\sqrt{z}\left\vert \underline{x}-\underline{x}%
_{k}\right\vert }}{4\pi\left\vert \underline{x}-\underline{x}_{k}\right\vert
}\right\Vert _{L^{2}(\Omega)}+\left\vert \varepsilon\right\vert \left\Vert
\mathcal{G}_{0}^{-\lambda_{\bar{n}}+\varepsilon,k}-\sum_{m=1}^{M_{\bar{n}}%
}\frac{\phi_{\bar{n},m}^{\ast}(\underline{x}_{k})\,\phi_{\bar{n},m}%
}{\varepsilon}\right\Vert _{L^{2}(\Omega)}%
\]
The first contribution at the r.h.s. of this expression is clearly an
$O(\varepsilon)$ being $\frac{e^{-\sqrt{z}\left\vert \underline{x}%
-\underline{x}_{k}\right\vert }}{4\pi\left\vert \underline{x}-\underline
{x}_{k}\right\vert }\in L^{2}(\Omega)$. Concerning the second contribution, we
notice that $\mathcal{G}_{0}^{z,k}-\sum_{m=1}^{M_{n}}\frac{\phi_{\bar{n}%
,m}^{\ast}(\underline{x}_{k})\,\phi_{\bar{n},m}}{\lambda_{\bar{n}}+z}$ is the
value in $\underline{x}_{k}$ of the integral kernel of the operator $\left(
-\Delta+z\right)  ^{-1}$ restricted to the space $L^{2}(\Omega)\backslash
l.s\left\{  \phi_{\bar{n},m}\right\}  _{m=1}^{M_{\bar{n}}}$. It belongs to
$L^{2}(\Omega)\backslash l.s\left\{  \phi_{\bar{n},m}\right\}  _{m=1}%
^{M_{\bar{n}}}$ for all $\underline{x}_{k}\in\Omega$ and $\,z\in
\mathbb{C}\backslash\left\{  -\lambda_{n}\right\}  _{\substack{n\in
\mathbb{N}\\n\neq\bar{n}}}$, it follows that%
\[
\lim_{z\rightarrow\lambda_{\bar{n}}}\left\Vert \mathcal{G}_{0}^{z,k}%
-\sum_{m=1}^{M_{n}}\frac{\phi_{\bar{n},m}^{\ast}(\underline{x}_{k}%
)\,\phi_{\bar{n},m}}{\lambda_{\bar{n}}+z}\right\Vert _{L^{2}(\Omega)}<\infty
\]
which implies%
\[
\lim_{\left\vert \varepsilon\right\vert \rightarrow0}\left\vert \varepsilon
\right\vert \left\Vert \mathcal{G}_{0}^{-\lambda_{\bar{n}}+\varepsilon,k}%
-\sum_{m=1}^{M_{\bar{n}}}\frac{\phi_{\bar{n},m}^{\ast}(\underline{x}%
_{k})\,\phi_{\bar{n},m}}{\varepsilon}\right\Vert _{L^{2}(\Omega)}=0
\]

\end{proof}

Making use of the previous Lemma, we study the behavior of $R_{z}^{AB}\varphi
$, $\varphi\in L^{2}(\Omega)$, as $z$ approaches non-degenerate spectral
points of $-\Delta$.

\begin{theorem}
\label{singularities}Let $\lambda_{\bar{n}}$ be a non-degenerate eigenvalue of
the operator $-\Delta$ defined on $H^{2}\cap H_{0}^{1}(\Omega)$, and assume
the corresponding eigenfunction $\phi_{\bar{n}}$ satisfies the following
condition%
\[
\phi_{\bar{n}}(\underline{x}_{k})\neq\underline{0}\in\mathbb{C}^{N}%
\]
then for any $\varphi\in L^{2}(\Omega)$, the limit $\lim_{z\rightarrow
-\lambda_{\bar{n}}}R_{z}^{AB}\varphi$ belongs to $L^{2}(\Omega)$.
\end{theorem}

\begin{proof}
Our aim is to prove that the point $-\lambda_{\bar{n}}$ is not a singular
point for $R_{z}^{AB}$. From the explicit expression of $R_{z}\varphi$ (see
the representation (\ref{Res free}) above), it follows that%
\begin{equation}
R_{z}^{AB}\varphi=\sum_{n\in\mathbb{N}}\sum_{m=1}^{M_{n}}\frac{\left(
\varphi,\phi_{n,m}\right)  }{\lambda_{n}+z}\phi_{n,m}+\sum_{n\in\mathbb{N}%
}\sum_{m=1}^{M_{n}}\frac{\left(  \varphi,\phi_{n,m}\right)  }{\lambda_{n}%
+z}\sum_{j,k=1}^{N}\left(  \Gamma(z)+A\right)  _{jk}^{-1}\,\phi_{n,m}%
(\underline{x}_{k})\,\mathcal{G}_{0}^{z,j} \label{R_AB 2}%
\end{equation}
Set $z=-\lambda_{\bar{n}}+\varepsilon$, $\varepsilon\in\mathbb{C}$; we shall
denote with $\left[  R_{-\lambda_{\bar{n}}+\varepsilon}^{AB}\varphi\right]
_{s}$ the singular part of $R_{-\lambda_{\bar{n}}+\varepsilon}^{AB}\varphi$
for $\left\vert \varepsilon\right\vert \rightarrow0$; this function is given
by%
\begin{equation}
\left[  R_{-\lambda_{\bar{n}}+\varepsilon}^{AB}\varphi\right]  _{s}%
=\frac{\left(  \varphi,\phi_{\bar{n}}\right)  }{\varepsilon}\,\phi_{\bar{n}%
}+\frac{\left(  \varphi,\phi_{\bar{n}}\right)  }{\varepsilon}\sum_{j,k=1}%
^{N}\left(  \Gamma(-\lambda_{\bar{n}}+\varepsilon)+A\right)  _{jk}^{-1}%
\,\phi_{\bar{n}}(\underline{x}_{k})\,\mathcal{G}_{0}^{-\lambda_{\bar{n}%
}+\varepsilon,\,j} \label{R_AB singular}%
\end{equation}
Replacing $\mathcal{G}_{0}^{-\lambda_{\bar{n}}+\varepsilon,\,j}$ with
(\ref{G_0(z) def}), from the previous expression we get%
\begin{align}
\left[  R_{-\lambda_{\bar{n}}+\varepsilon}^{AB}\varphi\right]  _{s}  &
=\left(  \varphi,\phi_{\bar{n}}\right)  \sum_{j,k=1}^{N}\frac{\left(
\Gamma(-\lambda_{\bar{n}}+\varepsilon)+A\right)  _{jk}^{-1}}{\varepsilon
}\,\phi_{\bar{n}}(\underline{x}_{k})\,\frac{e^{-\sqrt{-\lambda_{\bar{n}%
}+\varepsilon}\left\vert \cdot-\underline{x}_{j}\right\vert }}{4\pi\left\vert
\cdot-\underline{x}_{j}\right\vert }+\nonumber\\
&  +\frac{\left(  \varphi,\phi_{\bar{n}}\right)  }{\varepsilon}\left[
\phi_{\bar{n}}-\sum_{j,k=1}^{N}\left(  \Gamma(-\lambda_{\bar{n}}%
+\varepsilon)+A\right)  _{jk}^{-1}\,\phi_{\bar{n}}(\underline{x}%
_{k})\,h_{-\lambda_{\bar{n}}+\varepsilon,\,j}\right]  \label{R_AB 2.1}%
\end{align}
We will study the two contributions in (\ref{R_AB 2.1}) separately.

\begin{enumerate}
\item Define the vector $u_{j}^{\varepsilon}\in\mathbb{C}^{N}$ in the
following way%
\begin{equation}
\phi_{\bar{n}}(\underline{x}_{k})=\sum_{j}\varepsilon\left(  \Gamma
(-\lambda_{\bar{n}}+\varepsilon)+A\right)  _{kj}\,u_{j}^{\varepsilon}
\label{u_eps,j 1}%
\end{equation}
Using the relation (\ref{h_(z,k)_lim}) with $M_{\bar{n}}=1$, we have%
\begin{equation}
\lim_{\left\vert \varepsilon\right\vert \rightarrow0}\varepsilon\left(
\Gamma(-\lambda_{\bar{n}}+\varepsilon)+A\right)  _{kj}=\left\{
\begin{array}
[c]{l}%
-\phi_{\bar{n}}^{\ast}(\underline{x}_{j})\phi_{\bar{n}}(\underline{x}%
_{k})\qquad j\neq k\\
-\left\vert \phi_{\bar{n}}(\underline{x}_{k})\right\vert ^{2}\qquad j=k
\end{array}
\right.  \label{Gamma_kj lim}%
\end{equation}
then, limit of (\ref{u_eps,j 1}) as $\left\vert \varepsilon\right\vert
\rightarrow0$ is given by%
\begin{equation}
\phi_{\bar{n}}(\underline{x}_{k})=-\phi_{\bar{n}}(\underline{x}_{k})\sum
_{j}\phi_{\bar{n},}^{\ast}(\underline{x}_{j})\,u_{j}^{0} \label{u_eps,j 2}%
\end{equation}
from which it follows%
\begin{equation}
-\sum_{j}\phi_{\bar{n},m^{\prime}}^{\ast}(\underline{x}_{j})\,u_{j,m}(0)=1
\label{u_eps,j 3}%
\end{equation}
Under the assumption $\phi_{\bar{n}}(\underline{x}_{k})\neq\underline{0}$,
condition (\ref{u_eps,j 3}) defines a subset of vectors $u_{j}^{0}%
\in\mathbb{C}^{N}$ whose scalar product with $\phi_{\bar{n}}(\underline{x}%
_{j})$ is equal to $1$. We conclude that, in the limit $\left\vert
\varepsilon\right\vert \rightarrow0$, the first contribution at the r.h.s. of
(\ref{R_AB 2.1}) is given by $\left(  \varphi,\phi_{\bar{n}}\right)
\sum_{j=1}^{N}u_{j}^{0}\,\frac{e^{-\sqrt{-\lambda_{\bar{n}}+\varepsilon
}\left\vert \cdot-\underline{x}_{j}\right\vert }}{4\pi\left\vert
\cdot-\underline{x}_{j}\right\vert }$ which is clearly in $L^{2}(\Omega)$.

\item We use definition (\ref{u_eps,j 1}), in order to write the second term
at the r.h.s. of (\ref{R_AB 2.1}) as%
\begin{equation}
\frac{\left(  \varphi,\phi_{\bar{n}}\right)  }{\varepsilon}\left[  \phi
_{\bar{n}}-\sum_{j=1}^{N}u_{j}^{\varepsilon}\,\varepsilon h_{-\lambda_{\bar
{n}}+\varepsilon,\,j}\right]  \label{R_AB 3}%
\end{equation}
From (\ref{h_(z,k)_lim}) and the previous result (\ref{u_eps,j 3}), it follows%
\begin{equation}
\lim_{\left\vert \varepsilon\right\vert \rightarrow0}\left[  \phi_{\bar{n}%
}-\sum_{j=1}^{N}u_{j}^{\varepsilon}\,\varepsilon h_{-\lambda_{\bar{n}%
}+\varepsilon,\,j}\right]  =0
\end{equation}
In order to study the behavior of (\ref{R_AB 3}) as $\left\vert \varepsilon
\right\vert \rightarrow0$, we need to define the derivative $\frac
{d}{d\varepsilon}u_{j}^{\varepsilon}$ for $\varepsilon=0$; to this aim we
notice that from relation (\ref{u_eps,j 1}) it follows%
\begin{equation}
\sum_{j}\varepsilon\left(  \Gamma(-\lambda_{\bar{n}}+\varepsilon)+A\right)
_{kj}\,\frac{d}{d\varepsilon}u_{j}^{\varepsilon}=-\sum_{j}\frac{d}%
{d\varepsilon}\left[  \varepsilon\left(  \Gamma(-\lambda_{\bar{n}}%
+\varepsilon)+A\right)  _{kj}\right]  \,u_{j}^{\varepsilon} \label{u_eps,j 4}%
\end{equation}
Being%
\begin{equation}
\left.  \frac{d}{d\varepsilon}\left[  \varepsilon\left(  \Gamma(-\lambda
_{\bar{n}}+\varepsilon)+A\right)  _{kj}\right]  \right\vert _{\varepsilon
=0}=\left\{
\begin{array}
[c]{l}%
\lim_{\left\vert \varepsilon\right\vert \rightarrow0}\frac{d}{d\varepsilon
}\varepsilon h_{z,j}(\underline{x}_{k})\qquad j\neq k\\
\alpha_{k}+\lim_{\left\vert \varepsilon\right\vert \rightarrow0}\frac
{d}{d\varepsilon}\varepsilon h_{z,k}(\underline{x}_{k})\qquad j=k
\end{array}
\right.  \label{Gamma'_kj lim}%
\end{equation}
and using (\ref{Gamma_kj lim}), we obtain%
\begin{equation}
-\sum_{j}\phi_{\bar{n}}^{\ast}(\underline{x}_{j})\phi_{\bar{n}}(\underline
{x}_{k})\,\left.  \frac{d}{d\varepsilon}u_{j}^{\varepsilon}\right\vert
_{\varepsilon=0}=-\sum_{j}\left(  \lim_{\left\vert \varepsilon\right\vert
\rightarrow0}\frac{d}{d\varepsilon}\varepsilon h_{z,j}(\underline{x}%
_{k})+\alpha_{k}\delta_{jk}\right)  u_{j}^{0} \label{u_eps,j 5}%
\end{equation}
Due to the arbitrariness in the choice of $\underline{x}_{k}$, we may intend
the previous relation in a local sense%
\begin{equation}
-\sum_{j}\phi_{\bar{n}}^{\ast}(\underline{x}_{j})\phi_{\bar{n}}(\underline
{x})\,\left.  \frac{d}{d\varepsilon}u_{j}^{\varepsilon}\left(  \underline
{x}\right)  \right\vert _{\varepsilon=0}=-\sum_{j}\left(  \lim_{\left\vert
\varepsilon\right\vert \rightarrow0}\frac{d}{d\varepsilon}\varepsilon
h_{z,j}\left(  \underline{x}\right)  +\alpha_{k}\delta_{jk}\right)  u_{j}%
^{0}\left(  \underline{x}\right)  \label{u_eps,j 6}%
\end{equation}
where $\underline{x}\in\left\{  \underline{x}_{k}\right\}  _{k=1}^{N}$ and
$u_{j}^{\varepsilon}\left(  \underline{x}\right)  \in L^{2}(\Omega)$ is
defined by (\ref{u_eps,j 1}). Next we evaluate the limit $\left\vert
\varepsilon\right\vert \rightarrow0$ of the function (\ref{R_AB 3})%
\begin{gather*}
\lim_{\left\vert \varepsilon\right\vert \rightarrow0}\frac{\left(
\varphi,\phi_{\bar{n}}\right)  }{\varepsilon}\left[  \phi_{\bar{n}}-\sum
_{j=1}^{N}u_{j}^{\varepsilon}\,\varepsilon h_{-\lambda_{\bar{n}}%
+\varepsilon,\,j}\right]  =-\left(  \varphi,\phi_{\bar{n}}\right)
\lim_{\left\vert \varepsilon\right\vert \rightarrow0}\frac{d}{d\varepsilon
}\sum_{j=1}^{N}u_{j}^{\varepsilon}\,\varepsilon h_{-\lambda_{\bar{n}%
}+\varepsilon,\,j}=\\
=\left(  \varphi,\phi_{\bar{n}}\right)  \left(  \sum_{j=1}^{N}\phi_{\bar{n}%
}^{\ast}(\underline{x}_{j})\phi_{\bar{n}}\,\lim_{\left\vert \varepsilon
\right\vert \rightarrow0}\frac{d}{d\varepsilon}u_{j}^{\varepsilon}-\sum
_{j=1}^{N}u_{j}^{\varepsilon}\,\lim_{\left\vert \varepsilon\right\vert
\rightarrow0}\frac{d}{d\varepsilon}\varepsilon h_{-\lambda_{\bar{n}%
}+\varepsilon,\,j}\right)
\end{gather*}
Exploiting (\ref{u_eps,j 6}), we obtain an $L^{2}(\Omega)$ function%
\begin{equation}
\lim_{\left\vert \varepsilon\right\vert \rightarrow0}\frac{\left(
\varphi,\phi_{\bar{n}}\right)  }{\varepsilon}\left[  \phi_{\bar{n}}-\sum
_{j=1}^{N}u_{j}^{\varepsilon}\,\varepsilon h_{-\lambda_{\bar{n}}%
+\varepsilon,\,j}\right]  =\left(  \varphi,\phi_{\bar{n}}\right)  \sum
_{j}\alpha_{j}\,u_{j}^{0} \label{R_AB 3 lim}%
\end{equation}

\end{enumerate}
\end{proof}

\section{One point interaction in the ball: an exact solvable model}

In this section we shall study the selfadjoint extensions of $H_{0}$ when
$\Omega$ is a ball of radius $R$, centered in the origin of $\mathbb{R}^{3}$.
With the notation of the previous section, we will consider the case%
\begin{equation}
N=1,\ \underline{x}_{1}=\underline{0},\ A=\alpha\in\mathbb{R},\ B=1
\label{parameters}%
\end{equation}
With this choice of parameters, the operator $H^{AB}$ describes a point
interaction acting in the origin of $\Omega$; in what follows it will be
denoted as $H_{\alpha}$.

In order to characterize the spectral properties of $H_{\alpha}$, we
preliminary recall some basic facts about the Laplacian operator in the ball.
Set $\mathbb{N}_{0}\equiv\mathbb{N\cup}\left\{  0\right\}  $ and let $\left\{
\lambda_{n,k}\right\}  _{n\in\mathbb{N}_{0},\,k\in\mathbb{N}}$ be the point
spectrum of $-\Delta$ in the ball $\Omega$ with homogeneous Dirichlet
conditions boundary: it is well known that, for fixed $n$, the $\lambda_{n,k}$
are positive and $2n+1$ times degenerate; the corresponding eigenfuctions
$\left\{  \psi_{n,m,k},\ n\in\mathbb{N}_{0},\ m\in\mathbb{Z\cap}\left[
-n,n\right]  ,\ k\in\mathbb{N}\right\}  $ form a basis for the space
$L^{2}(\Omega)$. A general expression of $\psi_{n,m,k}$ in polar coordinates
is%
\begin{equation}
\psi_{n,m,k}(r,\vartheta,\varphi)=j_{n}(\sqrt{\lambda_{n,k}}\,r)\,Y_{n,m}%
(\vartheta,\varphi) \label{stati stazionari 1}%
\end{equation}
where the radial part is given in terms of the spherical Bessel functions of
order $n$, $j_{n}$, and the angular part is a combination of spherical
harmonics. For any $n\in\mathbb{N}_{0}$, \ the values $\lambda_{n,k}$ are
fixed by the boundary condition%
\begin{equation}
j_{n}(\sqrt{\lambda_{n,k}}\,R)=0 \label{stati stazionari 2}%
\end{equation}

From relations (\ref{H_AB_dom})-(\ref{Gamma_kj}) in Theorem \ref{Theorem H_AB}
and form assumptions (\ref{parameters}) we get the following representation%
\begin{gather}
D(H_{\alpha})=\left\{  \psi\in L^{2}(\Omega)\right.  \left\vert \,\psi
=\phi^{\lambda}+q\mathcal{G}_{0}^{\lambda}\right.  ,\ \phi^{\lambda}\in
H^{2}\cap H_{0}^{1}(\Omega),\nonumber\\
\qquad\qquad\qquad\phi^{\lambda}(\underline{0})=q\left(  \alpha+\frac
{\sqrt{\lambda}}{4\pi}+h_{\lambda}(\underline{0})\right)  ,\ \left.
\lambda\in\mathbb{C}\backslash\left\{  \lambda_{n,k}\right\}  _{n\in
\mathbb{N}_{0},\,k\in\mathbb{N}}\right\}  \label{H_alpha dom}%
\end{gather}%
\begin{equation}
H_{\alpha}\psi=-\Delta\phi^{\lambda}-\lambda q\mathcal{G}_{0}^{\lambda}
\label{H_alpha}%
\end{equation}
\begin{equation}
R_{z}^{\alpha}\varphi=R_{z}\varphi+\frac{R_{z}\varphi(\underline{0})}%
{\alpha+\frac{\sqrt{z}}{4\pi}+h_{z}(\underline{0})}\mathcal{G}_{0}^{z}
\label{R_alpha 1}%
\end{equation}
with%
\begin{equation}
\mathcal{G}_{0}^{z}(\underline{x})=\frac{e^{-\sqrt{z}\left\vert \underline
{x}\right\vert }}{4\pi\left\vert \underline{x}\right\vert }-h_{z}%
(\underline{x}) \label{G_0(z) 1 def}%
\end{equation}%
\begin{equation}
\left\{
\begin{array}
[c]{l}%
\medskip\left(  -\Delta+z\right)  h_{z}=0\\
\left.  h_{z}\right\vert _{\partial\Omega}=\frac{e^{-\sqrt{z}R}}{4\pi R}%
\end{array}
\right.  \label{h_(z,k) 1}%
\end{equation}

As it was stressed in the previous section, the action of operator $R_{z}$ can
be expressed by means of the eigenfuctions of the Dirichlet Laplacian in
$\Omega$%
\begin{equation}
R_{z}\varphi=\sum_{\substack{n\in\mathbb{N}_{0}\\m\in\mathbb{Z\cap}\left[
-n,n\right]  \\k\in\mathbb{N}}}\frac{\left(  \varphi,\psi_{n,m,k}\right)
}{\lambda_{n,k}+z}\psi_{n,m,k} \label{Res free 1}%
\end{equation}
Replacing (\ref{Res free 1}) into (\ref{R_alpha 1}) we get%
\begin{equation}
R_{z}^{\alpha}\varphi=\sum_{\substack{n\in\mathbb{N}_{0}\\m\in\mathbb{Z\cap
}\left[  -n,n\right]  \\k\in\mathbb{N}}}\frac{\left(  \varphi,\psi
_{n,m,k}\right)  }{\lambda_{n,k}+z}\psi_{n,m,k}+\sum_{\substack{n\in
\mathbb{N}_{0}\\m\in\mathbb{Z\cap}\left[  -n,n\right]  \\k\in\mathbb{N}}%
}\frac{\left(  \varphi,\psi_{n,m,k}\right)  }{\lambda_{n,k}+z}\frac
{\psi_{n,m,k}(\underline{0})}{\alpha+\frac{\sqrt{z}}{4\pi}+h_{z}(\underline
{0})}\mathcal{G}_{0}^{z} \label{R_alpha 2}%
\end{equation}
The explicit knowledge of the system $\left\{  \psi_{n,m,k}\right\}  $, will
allow a direct calculation of the spectral properties of $H_{\alpha}$. The
Fourier expansion of $\mathcal{G}_{0}^{z}$ in terms of the vectors $\psi
_{n,k}$ is: $\mathcal{G}_{0}^{z}=\sum_{_{n,m,k}}\frac{\psi_{n,m,k}^{\ast
}(\underline{0})}{\lambda_{n,k}+z}\psi_{n,m,k}$; the relation%
\begin{equation}
j_{n}(0)=\left\{
\begin{array}
[c]{c}%
0\quad n\neq0\\
1\quad n=0
\end{array}
\right.  \label{stati stazionari 3}%
\end{equation}
implies: $\psi_{n,m,k}^{\ast}(\underline{0})=\delta_{n\,0}$, and a simplified
expression of $\mathcal{G}_{0}^{z}$ follows%
\begin{equation}
\mathcal{G}_{0}^{z}=\sum_{_{k\in\mathbb{N}}}\frac{\psi_{0,0,k}}{\lambda
_{0,k}+z} \label{G_0(z) 2}%
\end{equation}
From (\ref{G_0(z) 2}) one can deduce an explicit form of the solution of
(\ref{h_(z,k) 1}). Infact, using the definition (\ref{G_0(z) 1 def}), we have%
\begin{equation}
h_{z}(r,\vartheta,\varphi)=\frac{e^{-\sqrt{z}r}}{4\pi r}-\sum_{_{k\in
\mathbb{N}}}\frac{\psi_{0,0,k}(r)}{\lambda_{0,k}+z} \label{h_(z,k) 2}%
\end{equation}
Moreover, as a direct calculation shows, for any $z\in\mathbb{C}%
\backslash\left\{  -\lambda_{n,k}\right\}  _{n\in\mathbb{N}_{0},\,k\in
\mathbb{N}}$ the expansion of the function $\frac{e^{-\sqrt{z}r}}{4\pi r}$
w.r.t. the system $\left\{  \psi_{n,m,k}\right\}  $ is%
\begin{equation}
\frac{e^{-\sqrt{z}r}}{4\pi r}=\sum_{_{k\in\mathbb{N}}}\frac{\psi_{0,0,k}%
^{\ast}(\underline{0})}{\lambda_{0,k}+z}\psi_{0,0,k}(r)-\frac{e^{-\sqrt{z}R}%
}{4\pi R}\,\frac{j_{0}(\sqrt{-z}\,r)}{j_{0}(\sqrt{-z}\,R)} \label{h_(z,k) 3}%
\end{equation}
Replacing this expression in (\ref{h_(z,k) 2}), we get%
\begin{equation}
\forall z\in\mathbb{C}\backslash\left\{  -\lambda_{n,k}\right\}
_{n\in\mathbb{N}_{0},\,k\in\mathbb{N}}\longrightarrow h_{z}(r,\vartheta
,\varphi)=-\frac{e^{-\sqrt{z}R}}{4\pi R}\,\frac{j_{0}(\sqrt{-z}\,r)}%
{j_{0}(\sqrt{-z}\,R)} \label{h_(z,k) 3.1}%
\end{equation}

The following theorem characterize the spectrum of $H_{\alpha}$.

\begin{theorem}
\label{Theorem_spectrum}Let $H_{\alpha}$ be defined by (\ref{H_AB_dom}%
)-(\ref{H_AB}) and (\ref{parameters}) and denote with $\sigma_{\alpha}$ its
spectrum. For any $\alpha\in\mathbb{R}\backslash\left(  0,-4\pi R\right]  $ we
have
\begin{equation}
\sigma_{\alpha}=\left\{  \lambda_{n,k}\right\}  _{n,k\in\mathbb{N}}%
\cup\left\{  \zeta_{n}\right\}  _{\substack{n\in\mathbb{N}\\n\ odd}}
\label{spettro 1}%
\end{equation}
with $\zeta_{n}\in\mathbb{R}^{+}$, $\left\{  \lambda_{n,k}\right\}
_{n,k\in\mathbb{N}}\cap\left\{  \zeta_{n}\right\}  _{\substack{n\in
\mathbb{N}\\n\ odd}}=\varnothing$ and%
\begin{equation}
\lim_{\substack{n\rightarrow\infty\\n\ odd}}\left(  \sqrt{\zeta_{n}}%
-\frac{n\pi}{2R}\right)  =0 \label{autovalori}%
\end{equation}
For $\alpha\in\left(  0,-4\pi R\right]  $ the spectrum $\sigma_{\alpha}$
acquires a negative point%
\begin{equation}
\sigma_{\alpha}=\left\{  \lambda_{n,k}\right\}  _{n,k\in\mathbb{N}}%
\cup\left\{  \zeta_{n}\right\}  _{\substack{n\in\mathbb{N}\\n\ odd}%
}\cup\left\{  \zeta_{-1}\right\}  \label{spettro 2}%
\end{equation}
with $\zeta_{-1}<0$. In the case $\alpha=0$, the set $\sigma_{0}$ is given by%
\begin{equation}
\sigma_{0}=\left\{  \lambda_{n,k}\right\}  _{n,k\in\mathbb{N}}\cup\left\{
\eta_{n}\right\}  _{n\in\mathbb{N}} \label{spettro 0}%
\end{equation}
where $\eta_{n}\in\mathbb{R}^{+}$, $\left\{  \lambda_{n,k}\right\}
_{n,k\in\mathbb{N}}\cap\left\{  \eta_{n}\right\}  _{n\in\mathbb{N}}$ are the
solutions of%
\begin{equation}
\frac{\sqrt{\eta_{n}}\sin2\sqrt{\eta_{n}}R}{1-\cos2\sqrt{\eta_{n}}R}%
=0\quad\forall n \label{spettro 0_def}%
\end{equation}

\end{theorem}

\begin{proof}
Due to the definition $R_{z}^{\alpha}=\frac{1}{H_{\alpha}+z}$, the spectral
points of $H_{\alpha}$ coincide with the singularities of the operators
$R_{z}^{\alpha}$, w.r.t. the variable $z$, modulo a change of sign. From
(\ref{R_alpha 2}) we see that $R_{z}^{\alpha}\varphi$ can be ill defined both
for $z\in\left\{  -\lambda_{n,k}\right\}  _{n\in\mathbb{N}_{0},\,k\in
\mathbb{N}}$ as for the zeroes of the function%
\begin{equation}
f^{\alpha}(z)=\alpha+\frac{\sqrt{z}}{4\pi}+h_{z}(\underline{0})
\label{Eigenval}%
\end{equation}
moreover we notice that, as it has been shown in Lemma \ref{Lemma!},
$h_{z}(\underline{0})$ is singular in the limit $z\rightarrow-\lambda_{n,k}$;
then none of these points may coincide with a zero of $f^{\alpha}(z)$; this
circumstance allows us to consider the two cases separately. Concerning the
first case, $z\in\left\{  -\lambda_{n,k}\right\}  _{n\in\mathbb{N}_{0}%
,\,k\in\mathbb{N}}$, Theorem \ref{singularities} (with relation
(\ref{stati stazionari 3})) exclude the points $\left\{  -\lambda
_{0,k}\right\}  _{k\in\mathbb{N}}$ from the singularities of $R_{z}^{\alpha
}\varphi$, from which it follows%
\[
\left\{  -\lambda_{n,k}\right\}  _{n,\,k\in\mathbb{N}}\in\sigma_{\alpha}%
\]

In order to study the zeroes of (\ref{Eigenval}), we exploit definition
(\ref{h_(z,k) 3.1}) - for $z\in\mathbb{C}\backslash\left\{  -\lambda
_{n,k}\right\}  $ - and the properties of $j_{0}(x)=\frac{\sin x}{x}$, to
write $f^{\alpha}(z)$ in the form: $f^{\alpha}(z)=\alpha+\frac{\sqrt{z}}{4\pi
}+\frac{1}{4\pi}\frac{2\sqrt{z}}{e^{2\sqrt{z}R}-1}$; then the equation
$f^{\alpha}(z)=0$ reads as%
\begin{equation}
\alpha+\frac{\sqrt{z}}{4\pi}+\frac{1}{4\pi}\frac{2\sqrt{z}}{e^{2\sqrt{z}R}%
-1}=0 \label{eigenvalue}%
\end{equation}
Setting $z=-\xi$ in equation (\ref{eigenvalue}), we recover the equation for
the eigenvalues of operator $H_{\alpha}$%
\begin{equation}
\alpha+\frac{i\sqrt{\xi}}{4\pi}+\frac{1}{4\pi}\frac{2i\sqrt{\xi}}%
{e^{2i\sqrt{\xi}R}-1}=0 \label{Eigenval1}%
\end{equation}
The selfadjointness of $H_{\alpha}$, allows us to expect only real solutions
for (\ref{Eigenval1}). We need to distinguish between different situations:

\begin{enumerate}
\item $\alpha\neq0$; $\xi\geq0$. (\ref{Eigenval1}) can be written in the form:%
\begin{equation}
\frac{\sqrt{\xi}}{4\pi\,\alpha}=\frac{\cos2\sqrt{\xi}R-1}{\sin2\sqrt{\xi}R}
\label{Eigenval2}%
\end{equation}
The solutions of this equation are the intersection points between the two
curves corresponding to the right and left hand side of (\ref{Eigenval2}). In
Figure \ref{Fig 1}, we show their behavior, w.r.t. the variable $\sqrt{\xi}$,
when the conditions: $\alpha>0$, $4\pi\,\alpha=\frac{1}{5}$, and $R=1$ are
assumed%
\begin{center}
\includegraphics[
natheight=2.500200in,
natwidth=3.958300in,
height=2.5417in,
width=4.0075in
]%
{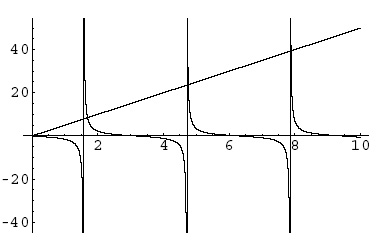}%
\\
Figure 1
\label{Fig 1}%
\end{center}

In both cases $\alpha>0$ and $\alpha<0$, the intersection points forms an
increasing sequence of non negative values $\left\{  \zeta_{n}\right\}
_{\substack{n\in\mathbb{N}\\n\ odd}}$ such that:%
\begin{equation}
\zeta_{1}=0 \label{Eigenval3}%
\end{equation}
and%
\begin{equation}
\lim_{\substack{n\rightarrow\infty\\n\ odd}}\left(  \sqrt{\zeta_{n}}%
-\frac{n\pi}{2R}\right)  =0 \label{Eigenval3.1}%
\end{equation}

\item $\alpha\neq0$; $\xi<0$. (\ref{Eigenval1}) can be written in the form%
\begin{equation}
\alpha+\frac{\sqrt{\left\vert \xi\right\vert }}{4\pi}+\frac{1}{4\pi}%
\frac{2\sqrt{\left\vert \xi\right\vert }}{e^{2\sqrt{\left\vert \xi\right\vert
}R}-1}=0\Rightarrow\frac{\sqrt{\left\vert \xi\right\vert }}{4\pi\,\alpha
}=\frac{1-e^{2\sqrt{\left\vert \xi\right\vert }R}}{1+e^{2\sqrt{\left\vert
\xi\right\vert }R}} \label{Eigenval4}%
\end{equation}
The solutions, defined by the intersection points of the two curves, are such
that%
\begin{equation}%
\begin{array}
[c]{ll}%
\alpha\geq-\left(  4\pi R\right)  ^{-1} & \text{no solution}\\
\alpha\in\left(  -\infty,-\left(  4\pi R\right)  ^{-1}\right)  &
\sqrt{\left\vert \zeta_{-1}\right\vert }%
\end{array}
\label{Eigenval5}%
\end{equation}
In Figure \ref{Fig 2} we show an example of the second case, $\alpha\in\left(
-\infty,-\left(  4\pi R\right)  ^{-1}\right)  $%
\begin{center}
\includegraphics[
natheight=2.322900in,
natwidth=3.999800in,
height=2.3627in,
width=4.0491in
]%
{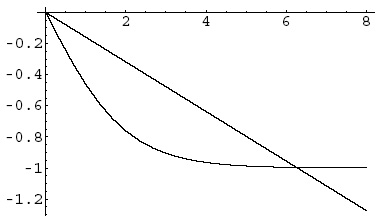}%
\\
Figure 2
\label{Fig 2}%
\end{center}

\item $\alpha=0$; $\xi>0$. Equation (\ref{Eigenval2}) becomes%
\begin{equation}
\frac{\sqrt{\xi}\sin2\sqrt{\xi}R}{1-\cos2\sqrt{\xi}R}=0 \label{Eigenval7}%
\end{equation}
which has infinitely many solutions; in Figure \ref{Fig 3}, the l.h.s. of
(\ref{Eigenval7}) is plotted w.r.t. the variable $\sqrt{\xi}$%
\begin{center}
\includegraphics[
natheight=2.479400in,
natwidth=3.947900in,
height=2.5201in,
width=3.9972in
]%
{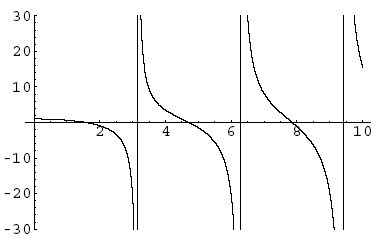}%
\\
Figure 3
\label{Fig 3}%
\end{center}

\item $\alpha=0$; $\xi<0$. Equation (\ref{Eigenval2}) becomes:%
\begin{equation}
\sqrt{\left\vert \xi\right\vert }\left(  1+\frac{2}{e^{2\sqrt{\left\vert
\xi\right\vert }R}-1}\right)  =0 \label{Eigenval8}%
\end{equation}

\end{enumerate}

which has no solution.
\end{proof}

The eigenfunctions of $H_{\alpha}$, related to the spectral sets
(\ref{spettro 1})-(\ref{spettro 0}), can be obtained directly from definition
(\ref{H_alpha}). Consider for instance the action of $H_{\alpha}$ on the
vectors $\psi_{n,m,k}$, $n\neq0$: by construction, these functions belong to
the operator domain and have null charge, from which it follows%
\begin{equation}
H_{\alpha}\psi_{n,m,k}=-\Delta\psi_{n,m,k}=\lambda_{n,k}\,\psi_{n,m,k}%
\qquad\forall n\neq0 \label{stati stazionari 4}%
\end{equation}
Then the set spanned by $\left\{  \psi_{n,m,k},\ m\in\mathbb{Z\cap}\left[
-n,n\right]  \right\}  $ represents the autospace of $H_{\alpha}$ related to
the eigenvalue $\lambda_{n,k}$.

On the other hand, if we consider a spectral point $\lambda\in\sigma_{\alpha
}\backslash\left\{  \lambda_{n,k}\right\}  _{n,k\in\mathbb{N}}$, from
(\ref{H_alpha}) it follows that the corresponding solution of the eigenvalue
problem is $q\mathcal{G}_{0}^{-\lambda}$ with $q\in\mathbb{C}$.

\section{Coupling one particle with a quantum bit}

In this section we briefly sketch how point interaction models can be used in
order to describe an information flow between two interacting quantum systems.
To this aim, making use of the results obtained above, we shall give a
complete characterization of the family of point interactions coupling a
quantum particle - confined in a finite volume of $\mathbb{R}^{3}$ - with a
two level quantum system (q-bit).

Let $\Omega$ be the ball of radius $R$ centered in the origin of
$\mathbb{R}^{3}$ and consider the symmetric operator%
\begin{equation}
\left\{
\begin{array}
[c]{l}%
\medskip T_{0}=H_{0}\otimes\mathbb{I}_{\mathbb{C}^{2}}+\mathbb{I}%
_{L^{2}(\Omega)}\otimes U\\
D(T_{0})=D(H_{0})\otimes\mathbb{C}^{2}%
\end{array}
\right.  ;\quad U=%
\begin{pmatrix}
E_{+} & 0\\
0 & E_{-}%
\end{pmatrix}
,\ E_{\pm}\in\mathbb{R} \label{T_0}%
\end{equation}
where $H_{0}$ is defined by (\ref{H_0}) while $U$ is the Hamiltonian
associated to the q-bit. The selfadjoint extensions of $T_{0}$ shall describe
zero range interactions, acting in the origin of the ball $\Omega$, between
the particle and the quantum bit. The construction of these extensions will be
performed along the same line followed in the previous Sections. From Lemma
\ref{Defect} follows that the defect spaces of $T_{0}$ are%
\[
\mathcal{H}_{+}=l.s.\left\{  \sum_{\sigma=\pm}\mathcal{G}_{0}^{-i-E_{\sigma}%
}\otimes\chi_{\sigma}\right\}  ;\quad\mathcal{H}_{-}=l.s.\left\{  \sum
_{\sigma=\pm}\mathcal{G}_{0}^{i+E_{\sigma}}\otimes\chi_{\sigma}\right\}
\]
where $\chi_{+}=\left(  1,0\right)  $, $\chi_{-}=\left(  0,1\right)  $ are the
eigenvectors of the matrix $A$, while the functions $\mathcal{G}_{0}^{z}$ are
defined by (\ref{G_0(z) 1 def})-(\ref{h_(z,k) 3.1}). Moreover, let $\Psi
=\sum_{\sigma=\pm}\psi_{\sigma}\otimes\chi_{\sigma}$, $\psi_{\sigma}\in
L^{2}(\Omega)$, be a generic vector in $D(T_{0}^{\ast})$; we define the maps
$\Lambda_{1,2}:D(T_{0}^{\ast})\rightarrow\mathbb{C}^{2}$ as follows%
\begin{equation}
\left(  \Lambda_{j}\Psi\right)  _{\sigma}=\Gamma_{j}\psi_{\sigma},\quad j=1,2
\label{Lambda}%
\end{equation}%
\begin{equation}
\Gamma_{1}\psi_{\sigma}=\lim_{\underline{x}\rightarrow\underline{0}}%
4\pi\left\vert \underline{x}\right\vert \psi(\underline{x}) \label{Lambda 1}%
\end{equation}%
\begin{equation}
\Gamma_{2}\psi_{\sigma}=\lim_{\underline{x}\rightarrow\underline{0}}\left(
\psi(\underline{x})-\frac{\Gamma_{1}\psi_{\sigma}}{4\pi\left\vert
\underline{x}\right\vert }\right)  \label{Lambda 2}%
\end{equation}
Proceeding as in Theorem \ref{BVT}, it is easy to show that $\left(
\mathbb{C}^{2},\Gamma_{1},\Gamma_{2}\right)  $ form a boundary value space for
the operator $T_{0}$. Proposition \ref{Panka} implies that all selfadjoint
extensions of $T_{0}$ are parametrized by couples of matrices $A,B\in$
$\mathbb{C}^{2,2}$ fulfilling the conditions $\left[  I\right]  $ and $\left[
II\right]  $ (see Section 2); each extension $T^{AB}$ is a restriction of the
adjoint $T_{0}^{\ast}$ to those vectors $\psi\in D(T_{0}^{\ast})$ satisfying
the relation%
\begin{equation}
A\Lambda_{1}\psi=B\Lambda_{2}\psi\label{s.a.b.c. 1}%
\end{equation}
The following Theorem provides an explicit representation of $T^{AB}$

\begin{theorem}
\label{Theorem T_AB}Let $A,B\in$ $\mathbb{C}^{2,2}$ satisfying the conditions
$\left[  I\right]  $ and $\left[  II\right]  $; for any $\lambda\in
\mathbb{C}\backslash\mathbb{R}$, the selfadjoint extension $T^{AB}$,
corresponding to the couple $A,B$, is described as follows%
\begin{align*}
D(T^{AB})  &  =\left\{  \Psi\right.  =\sum_{\sigma=\pm}\left(  \phi_{\sigma
}^{\lambda}+q_{\sigma}\mathcal{G}_{0}^{\lambda}\right)  \otimes\left.
\chi_{\sigma}\right\vert \ \phi_{\sigma}^{\lambda}\in H^{2}\cap H_{0}%
^{1}(\Omega);\\
q_{\sigma}  &  =\left(  \Lambda_{1}\Psi\right)  _{\sigma};\ \phi_{\sigma
}^{\lambda}(\underline{0})=\sum_{\sigma^{\prime}=\pm}\left[  \left(
h_{\lambda}(\underline{0})+\tfrac{\sqrt{\lambda}}{4\pi}\right)  \mathbb{I}%
+B^{-1}A\right]  _{\sigma\sigma^{\prime}}\left.  q_{\sigma^{\prime}}\right\}
\end{align*}%
\[
T^{AB}\Psi=\sum_{\sigma=\pm}\left(  -\Delta\phi_{\sigma}^{\lambda}-\lambda
q_{\sigma}\mathcal{G}_{0}^{\lambda}\right)  \otimes\chi_{\sigma}+\sum
_{\sigma=\pm}\left(  \phi_{\sigma}^{\lambda}+q_{\sigma}\mathcal{G}%
_{0}^{\lambda}\right)  \otimes U\chi_{\sigma}%
\]
Moreover given any $\Phi=\sum_{\sigma=\pm}\varphi_{\sigma}\otimes\chi_{\sigma
}$ in $L^{2}(\Omega)\otimes\mathbb{C}^{2}$, the action of the resolvent
operator $R_{z}^{AB}=\frac{1}{T^{AB}+z}$ on $\Phi$ is%
\[
R_{z}^{AB}\Phi=\sum_{\sigma=\pm}\left(  R_{z}\varphi_{\sigma}+\sum
_{\sigma^{\prime},\nu=\pm}\left(  B\left(  h_{\lambda}(\underline{0}%
)+\tfrac{\sqrt{\lambda}}{4\pi}\right)  \mathbb{I}+A\right)  _{\sigma
\sigma^{\prime}}^{-1}B_{\sigma^{\prime}\nu}R_{z}\varphi_{\nu}(\underline
{0})\,\mathcal{G}_{0}^{\lambda}\right)  \otimes\chi_{\sigma}%
\]

\end{theorem}

\begin{proof}
The proof is easily obtained following the same line of Theorem
\ref{Theorem H_AB}.
\end{proof}

In the perspective of applications, we consider of particular interest the
subfamily of selfadjoint extensions $T^{AB}$ described by the conditions%
\begin{equation}
A=%
\begin{pmatrix}
\alpha_{+} & 0\\
0 & \alpha_{-}%
\end{pmatrix}
;\quad B=\mathbb{I} \label{diag}%
\end{equation}
These operators, which will be denoted in the following as $T_{\left\{
\alpha_{+},\alpha_{-}\right\}  }$, are defined by%
\begin{equation}
D(T_{\left\{  \alpha_{+},\alpha_{-}\right\}  })=\left\{  \left.  \Psi
=\sum_{\sigma=\pm}\psi_{\sigma}\otimes\chi_{\sigma}\right\vert \ \psi_{\sigma
}\in D(H_{\alpha_{\sigma}})\right\}  \label{T_alpha_dom}%
\end{equation}%
\begin{equation}
T_{\left\{  \alpha_{+},\alpha_{-}\right\}  }\Psi=\sum_{\sigma=\pm}%
H_{\alpha_{\sigma}}\psi_{\sigma}\otimes\chi_{\sigma}+\sum_{\sigma=\pm}%
\psi_{\sigma}\otimes U\chi_{\sigma} \label{T_alpha}%
\end{equation}
where $H_{\alpha_{\sigma}}$ is an Hamiltonian with a point interaction acting
in the origin of the ball $\Omega$ with strength $\alpha_{\sigma}$ (see the
definitions (\ref{H_alpha dom})-(\ref{H_alpha}) and Theorem \ref{Theorem T_AB}).

Let us consider a quantum mechanical system consisting of a confined particle
interacting with a quantum bit through the Hamiltonian operator $T_{\left\{
\alpha_{+},\alpha_{-}\right\}  }$. The specific feature of this model rests
upon the fact that the particle may follow a different dynamics - generated
respectively by the Hamiltonians $H_{\alpha_{+}}$ and $H_{\alpha_{-}}$ -
depending on the state of the q-bit. At the same time, the latter evolves
freely, independently of the motion of the particle. Therefore, in the time
evolution of this system, we shall observe a flow of information from the
q-bit to the particle without energy exchanges. The extension of this model to
the case of $N$ confined quantum particles, the analysis of the implications
between the lack of coherence of the q-bit and the information flow to its
environment, as well as the construction of point interaction models of
quantum measurement apparata, will be considered in further work.

\end{document}